\documentclass[twocolumn,showpacs,preprintnumbers,amsmath,amssymb, prl]{revtex4}

\usepackage{graphicx}
\usepackage{dcolumn}
\usepackage{bm}
\usepackage{epsfig}

\begin{document}

\title{Possible Dynamic States in Inductively Coupled Intrinsic
Josephson Junctions of Layered High-$T_c$ Superconductors}

\author{Shizeng Lin\(^{1,2}\) and Xiao Hu\(^{1,2,3}\)}

\affiliation{\(^{1}\)WPI Center for Materials
Nanoarchitectonics, National Institute for Materials Science, Tsukuba 305-0047, Japan\\
\(^{2}\)Graduate School of Pure and Applied Sciences, University of
Tsukuba, Tsukuba 305-8571, Japan\\
\(^{3}\)Japan Science and Technology Agency, 4-1-8 Honcho,
Kawaguchi, Saitama 332-0012, Japan}
\date{\today}

\begin{abstract}
Based on computer simulations and theoretical analysis, a new
dynamic state is found in inductively coupled intrinsic Josephson
junctions in the absence of an external magnetic field. In this
state, the plasma oscillation is uniform along the $c$ axis with the
fundamental frequency satisfying the ac Josephson relation. There
are $(2m+1)\pi$ phase kinks around the junction center, with $m$
being an integer, periodic and thus non-uniform in the $c$
direction. In the IV characteristics, the state manifests itself as
current steps occurring at all cavity modes. Inside the current
steps, the plasma oscillation becomes strong, which generates
several harmonics in frequency spectra at a given voltage. The
recent experiments on terahertz radiations from the mesa of a BSCCO
single crystal can be explained in terms of this state.

\end{abstract}


\maketitle

Soon after the discovery of high-$T_c$ superconductivity in
cuprates, it was demonstrated that a single crystal of layered
superconductors, such as BSCCO, behaves as a stack of intrinsic
Josephson junctions (IJJs) \cite{Kleiner92}. The IJJs have been
investigated extensively since then, and various dynamical
properties were explored. One focus in this field is to exploit the
dynamic states of IJJs to generate electromagnetic waves in the
terahertz (THz) regime \cite{Koyama95, Ozyuzer07, kadowaki08}. The
high-$T_c$ cuprate superconductors are unique for this purpose since
the IJJs are homogeneous at atomic scale and the energy gap is
large, $60$meV, which make it possible to tune the frequency in a
wide range by adjusting the bias voltage. To obtain coherent
radiations, however, all the junctions should be operated in-phase,
which turns out to be not easy.

The dynamics of IJJs are appropriately described by the coupled
sine-Gorden equations \cite{Sakai93}
\begin{equation}\label{eq1}
\partial_x^2P_l=(1-\zeta \Delta^{(2)})(\sin P_l+\beta\partial_t P_l
+\partial_t^2P_l-J_{\rm{ext}}),
\end{equation}
where $P_l$ is the gauge-invariant phase difference at the $l$th
junction, $\beta\equiv4\pi\sigma_c\lambda_c/c\sqrt{\varepsilon_c}$
the normalized $c$-axis conductivity, $\zeta\equiv\lambda_{ab}^2/sD$
the inductive coupling and $\Delta^{(2)}P_l\equiv
P_{l+1}+P_{l-1}-2P_l$ the difference operator. $\varepsilon_c$ is
the dielectric constant and $\sigma_c$ is the conductance along the
$c$ axis. $\lambda_c$ and $\lambda_{ab}$ are the penetration depths.
$s$ ($D$) is the thickness of the superconducting (insulating)
layer. Length is normalized by $\lambda_c$, time is normalized by
the Josephson plasma frequency $1/2\pi f_J$ with $f_J\equiv
c/2\pi\lambda_c\sqrt{\varepsilon_c}$, and voltage is in unit of
$\Phi_0 f_J/c$. Here $c$ is the light velocity in vacuum.

Without an external in-plane magnetic field, Eq. (\ref{eq1}) have
the plasma solution, moving soliton and anti-soliton states, and
McCumber state \cite{Kleiner00}. The plasma solution corresponds to
the collective small phase oscillation in Eq. (\ref{eq1}). It
resonates in the cavity formed by IJJs when the cavity resonance
condition meets, i.e. $q\lambda_w/2=L$ with q being an integer,
$\lambda_w$ plasma wavelength and $L$ the length of the IJJs. It
corresponds to the voltage per layer $V=q\pi c_j/L$ with $c_j$ the
plasma velocity ($1\le j\le N$, $N$ the total number of junctions)
\cite{Sakai94} assuming that the ac Josephson relation is satisfied.

In the soliton state, a soliton (anti-soliton) moving with velocity
$v$ is reflected as an anti-soliton (soliton) by the edges. The
periodic motion and reflection of soliton and anti-soliton
contributes to a dc voltage $V=2\pi v/L$. When the velocity of
soliton approaches $c_j$, it gives birth to the zero-field step
(ZFS) at $V=2\pi c_j/L$ \cite{Fulton73, Krasnov99, Kleiner00}, which
corresponds to the $2$nd cavity mode. If $n$ solitons engage in this
process, they produce the $n$th ZFS. The interplay between solitons
and plasma leads to many peculiar behaviors \cite{Ustinov98}, such
as Cherenkov radiation \cite{Hechtfischer97}.

In the McCumber state, the phase is uniform through the $x$
direction. Therefore the junctions are decoupled and it gives linear
branches in the IV characteristics.

In the present study, we find that in the absence of an external
magnetic field, a new dynamic state can be achieved in the
inductively coupled sine-Gordon systems, which is characterized as
follows: in the IV characteristics it generates current steps at all
cavity modes; while the plasma oscillation is uniform along the $c$
axis, in the total phase there are static kinks of $(2m+1)\pi$ along
the $x$ axis, which are periodic but non-uniform in the $c$
direction. It is shown that the recent experiments
\cite{Ozyuzer07,kadowaki08} on terahertz emission can be understood
in terms of this new state.

\begin{figure}[t]
\setlength{\unitlength}{1cm}
 \psfig{figure=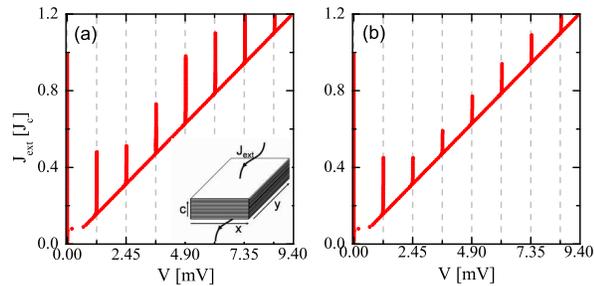,width=\columnwidth}
\caption{\label{f1}(color). Simulated IV characteristics under the
boundary conditions (a) without and (b) with radiation with
$C=0.000177$ and $R=707.1$. The vertical dashed lines correspond to
the voltage at cavity modes $V=n\Phi_0/(2\sqrt{\varepsilon_c}L)$.
Inset: geometry of the mesa of BSCCO single crystal. }
\end{figure}

The results shown below are for $\beta=0.02$, $\zeta=4/9\times
10^6$, typically for BSCCO, and $L=80\rm{\mu m}$; confirmations of
the main results have been made with other parameters. As displayed
in the inset of Fig. \ref{f1}, the current $J_{\rm ext}$ is fed in
along the $c$ axis. Since we are searching for states periodic in
the $c$ axis, we adopt the periodic boundary condition along this
direction which makes the maximum plasma velocity $c_0=1$ possible.
The equations of motion Eq. (\ref{eq1}) are integrated by the
staggered leapfrog algorithm. The time step in all simulations is
set to $\Delta t=0.0018$ and the mesh size is set to $\Delta
x=0.002$. The accuracy is checked with smaller $\Delta x$ and
$\Delta t$. We consider first the system under the boundary
condition $\partial_x P_l=0$, which turns out to be a good
unperturbed state for discussion on radiation as revealed later.

The important feature of the IV characteristics in Fig. \ref{f1}(a)
is the existence of current steps. In sharp contrast with the ZFS in
single junctions \cite{Fulton73}, the present current steps in IJJs
occur at both even and odd cavity modes; the dc voltage and the
frequency of plasma oscillation satisfy the ac Josephson relation.

\begin{figure}[b]
\psfig{figure=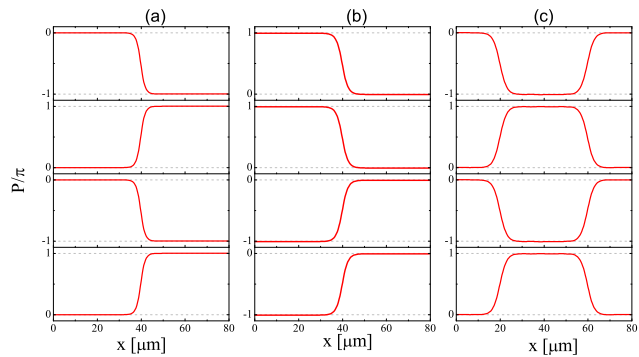,width=\columnwidth}
\caption{\label{f2}(color). Typical configurations of the static
term in the phase difference (the 2nd term in Eq. (\ref{eq2})): (a)
period of 2 layers and (b) period of 4 layers, corresponding to the
1st current step, and (c) corresponding to the 2nd current step. }
\end{figure}

As revealed by the present numerical simulations, there are four
parts in the phase variables: (1) a term evolving linearly with time
according to the ac Josephson relation $\omega t$, (2) a term of
$(2m+1)\pi$ phase kinks and anti-kinks arranged periodically in the
$c$ direction, (3) a plasma term uniform along the c axis
$\widetilde{P}(x,t)$, (4) a constant term $\varphi$. Figure \ref{f2}
exhibits several typical configurations for the 2nd term. For ease
of discussions, we restrict ourselves to the 1st cavity mode in the
following discussions, which can be extended to higher modes
straightforwardly. The total phase therefore can be expressed by
\begin{equation}\label{eq2}
P_l(x,t)=\omega t+f_{l}P_l^s(x)+\widetilde{P}(x,t)+\varphi,
\end{equation}
where $P_l^s(x)$ is a soliton from $0$ to $2\pi$ with the center at
$x=L/2$ (see below), $\widetilde{P}(x,t)=A\cos(\pi x/L)\sin \omega t
+$ high harmonics.

Substituting Eq. (\ref{eq2}) into Eq. (\ref{eq1}) and using the
expansion of sine of sine by the Bessel function, we arrive at the
following relation for the static parts of phases
\begin{equation}\label{eq3}
\begin{array}{l}
f_l\partial^2_xP_l^s = \omega\beta-J_{\rm
ext}\\-(1-\zeta\Delta^{(2)})J_1(A\cos(\pi
x/L))\sin(f_lP_l^s+\varphi),
 \end{array}
\end{equation}
where $J_1$ is the Bessel function of the first kind and for
simplicity only the fundamental mode is taken.

Off resonance the system behaves in an ohmic way as seen in Fig.
\ref{f1}, where $J_{\rm ext}\approx\beta\omega$ and
$\varphi\approx0$. The solutions of period of 2 layers (period-2),
where $f_l=(-1)^l f$ (Fig. \ref{f2}(a)),  and of 4 layers
(period-4), where $f_l=(-1)^{[l/2]} f$ (Fig. \ref{f2}(b)),
diagonalize the difference operator $\Delta^{(2)}$. In both cases,
$P_l^s$ is independent of $l$, thus $P_l^s=P^{s0}$. We then arrive
at the equation
\begin{equation}\label{eq4}
f\partial^2_xP^{s0} = -(1+q\zeta) J_1(A\cos(\pi x/L))\sin(f P^{s0}),
\end{equation}
where $q=4$ for period-2 and $q=2$ for period-4.

Since the l.h.s. of Eq.~(\ref{eq4}) and $J_1(A\cos(x\pi/L))$ are
antisymmetric with respect to $x=L/2$, $\sin(fP^{s0})$ must be
symmetric. This imposes $f=(2m+1)/2$, with an integer $m$.

When the voltage is tuned to one of the cavity modes, the plasma
oscillation can acquire large amplitude in the form of standing
waves. By taking a non-zero constant $\varphi$, the system can
bypass increasing $J_{\rm{ext}}$ through the Josephson current
\begin{equation}\label{eq5}
\begin{array}{l}
J_{\rm ext}=\beta\omega-\frac{\sin\varphi}{L}\int^L_0dx
J_1(A\cos(\pi x/L))\cos(f P^{s0}).
\end{array}
\end{equation}
This gives rise to the current steps in the IV characteristics in
Fig. \ref{f1}. Extra amount of energy besides the ohmic dissipation
is fed into the plasma oscillation.

\begin{figure}[t]
\setlength{\unitlength}{1cm}
 \psfig{figure=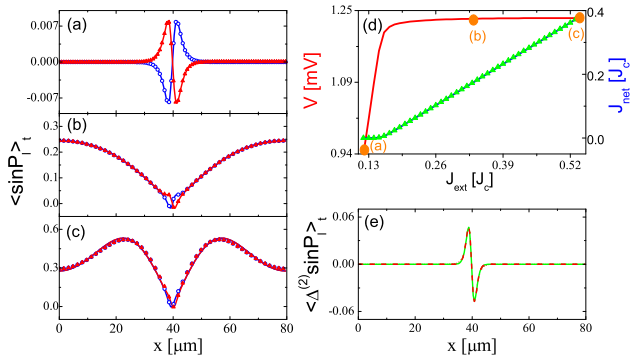,width=\columnwidth}
\caption{\label{f3}(color).  Comparison between the theory and
simulations on the dc part of Josephson current $\langle \sin
P_l\rangle_t$ in two neighboring layers for the configuration in
Fig. \ref{f2}(b): (a) off resonance at $J_{\rm{ext}}=0.12$, (b) at
the middle of the 1st step with $J_{\rm{ext}}=0.35$, (c) at the top
of the 1st step with $J_{\rm{ext}}=0.53$. The solid curves are
results by numerical simulations and the symbols are derived by the
analytic expressions. (d) Current dependence of voltage (red) and
$J_{\rm net}$ (blue) at the 1st current step. Green symbols are for
the theoretical estimate $J_{\rm ext}-\omega\beta$. (e)
$\langle\Delta^{(2)}\sin P_l\rangle_t$ at $J_{\rm{ext}}=0.35$
(dashed red line) and at $J_{\rm{ext}}=0.53$ (dashed green line). }
\end{figure}

In Figs.~\ref{f3} (a)-(c), we display distributions of the dc part
of Josephson current $\langle \sin P_l \rangle_t$ in two adjacent
junctions. The symbols are given by $J_1(A\cos(\pi x/L))\sin(\pm
P^{s0}(x)/2+\varphi)$ where
$P^{s0}(x)=4\arctan[\exp((x-L/2)/\lambda_J')]$ (see Eq.
(\ref{eq4})). The dependence of net Josephson current $J_{\rm
net}\equiv\langle \sin P_l\rangle_{x,t}$ on $J_{\rm{ext}}$ at the
1st current step is shown in Fig.~\ref{f3}(d).

The remaining terms in the r.h.s. of Eq. (\ref{eq3}), which are
proportional to the inductive coupling $\zeta$, should correspond to
a (unquantized) soliton as required by the l.h.s of the equation.
This can be seen clearly in Fig. \ref{f3}(e). The system arranges
itself in a way that the parameter $\zeta$ is blocked from involving
directly in the net Josephson current, which should be in order of
$J_c$.

\begin{figure}[b]
\psfig{figure=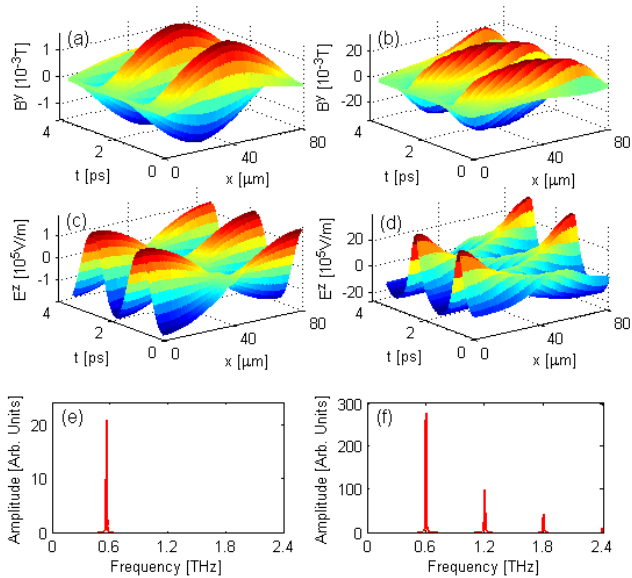,width=\columnwidth}
\caption{\label{f4}(color). Time evolutions of the magnetic field
and electric field and the frequency spectrum. The dc part of the
electric field has been subtracted. Left/right: at the bottom/top of
the 1st current step. }
\end{figure}

Despite that the total phase is \emph{non-uniform} along the $c$
axis, the electric field $E_l^z=\partial _t P_l$ as well as the
magnetic field $B_l^y=(1-\zeta\Delta^{(2)})^{-1}\partial_x P_l=
\partial_x \widetilde{P}+O(1/\zeta)$ with $P_l(x,t)$ given by Eq.
(\ref{eq2}) are \emph{uniform} along the $c$ axis.  This makes a
coherent superradiation possible as observed in the recent
experiments \cite{Ozyuzer07,kadowaki08}. Distributions of electric
and magnetic fields are displayed in Fig. \ref{f4}. In contrast with
the state at bottom of the current step (left of Fig. \ref{f4}), the
amplitude of plasma term increases and high harmonics become visible
inside the current steps, and thus the standing waves of
electromagnetic field are not simply trigonometric anymore (right of
Fig. \ref{f4}). The wave crest of magnetic field becomes smaller
when it approaches the edges and it changes sign when it is
reflected. It resembles the motion of solitons in ZFS except that
the magnetic flux is not quantized since it is associated with the
plasma oscillations.

From the foregoing minimum theoretical analysis, we have already
captured the main features of the system revealed by simulations.
The following discussions are necessary to make the picture more
comprehensive: (I) There are infinite number of configurations of
the static phase even in the same current step. Besides those in
Figs. \ref{f2}(a) and (b), periodic configurations of $(2m+1)\pi$
kinks, and their composites are observed in simulations.
Configurations of phase kinks not periodic in the $c$ axis may be
possible. Superpositions of soliton(s) and anti-soliton(s) along the
junction direction is also permitted as for the single sine-Gordon
system \cite{McLaughlin78}. The height of current steps depends on
the specific configuration as can be read from Eq. (\ref{eq5}). (II)
The phase configurations for higher cavity modes can be constructed
from the first cavity mode, see Fig. \ref{f2}(d). (III) In addition
to the plasma oscillation, there exists a small oscillation of the
center of soliton $P_l^s(x)$ as well, which produces a spike-like
oscillation with characteristic length $\lambda'_J$ in
electromagnetic fields. (IV) As revealed in the above discussions,
the strong inductive coupling is essential for the new dynamic
states, and therefore it is impossible to build such current steps
in a single Josephson junction. In fact, the current steps
completely vanish when $\zeta$ is small in simulations. (V)While
BSCCO may be the best choice, the new state can be realized in other
layered superconductors, including s-wave ones with not so high
$T_c$. The crystal is to be thick to reduce the surface effect.

Let us investigate the possible radiation of energy from the IJJs.
The boundary condition should be formulated explicitly for this
purpose. Theoretical efforts have been attempted to model the
interaction between IJJs and the outside space
\cite{Gronbechjensen89, Soriano96, Bulaevskii06PRL,Koshelev08PRB}.
We model the external load by an effective RC circuit
\cite{Soriano96}, and hence the boundary condition becomes
$\partial_t\widetilde{P}/\partial_x\widetilde{P}
=\widetilde{E}^z/\widetilde{B}^y=Z=R-i/(C\omega)=|Z|\exp(i\theta)$.
The sharp radiation peaks upon voltage sweeping observed in recent
experiments \cite{Ozyuzer07,kadowaki08} indicate that there is a
significant mismatch in impedance between IJJs and the outside
space, and that electromagnetic waves are reflected mostly at the
edges of IJJs. The small ratio between the mesa thickness in the
experiments and $\lambda_{c}$ may be a source of this mismatch
\cite{Koshelev08PRB}, which implies $|Z|\gg 1$. We have confirmed
that the new states remain stable for large values of $|Z|$.

The IV characteristics under the boundary condition permitting
radiation are shown in Fig. \ref{f1}(b). The heights of the current
steps are reduced when radiations are present, since when the
radiation is strong it is hard to form stable standing waves.

\begin{figure}[t]
\psfig{figure=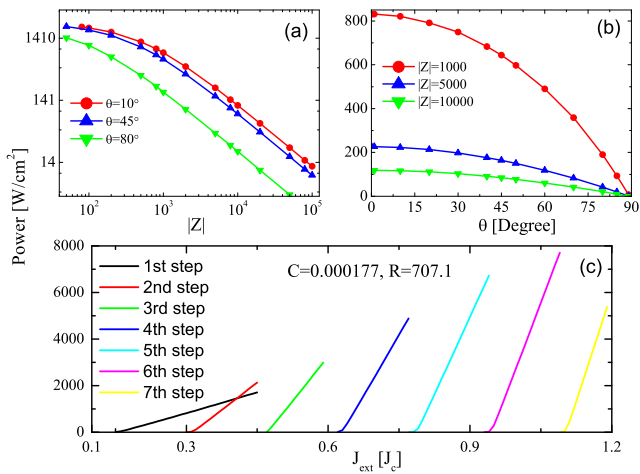,width=\columnwidth}
\caption{\label{f5}(color). Dependence of radiation power evaluated
by the Poynting vector at the edges of IJJs (a) on $|Z|$ and (b) on
$\theta$ inside the 1st current step at $J_{\rm{ext}}=0.27$. (c)
Radiation power at all current steps.}
\end{figure}

As shown in Fig. \ref{f5}(c), the radiation power increases linearly
with the external current at the current steps. This behavior can be
explained by the above theoretical description including
additionally the radiation as a small perturbation. The power at the
top of the 1st current step is of one order of magnitude stronger
than that at the bottom, and the maximum power with the present
boundary condition is about $8000\rm{W/cm^2}$, which is quite
promising for applications. The dependence of the radiation power on
$|Z|$ and $\theta$ is detailed in Figs. \ref{f5}(a) and (b), which
can be described as $\cos\theta/|Z|$.

The above simulation results seem to be able to provide a consistent
picture for the two recent experiments \cite{Ozyuzer07, kadowaki08}
on the terahertz radiation in zero external magnetic field. Coherent
radiations were detected in the resistive curve in Ref.
\cite{Ozyuzer07} with the frequency corresponding to the fundamental
mode of the 1st current step. In Ref. \cite{kadowaki08}, radiations
were detected in the anomalous region of IV curve, and the detected
power is much stronger than that in Ref. \cite{Ozyuzer07}. Comparing
these features with our simulated results, it is natural to assign
the radiations in Ref. \cite{kadowaki08} to states inside the 1st
current step, while those in Ref. \cite{Ozyuzer07} to states at the
bottom of the current step. The shape of the anomalous IV curve
where radiations are observed in Ref. \cite{kadowaki08} is also
indicative that the system falls into the current step. High cavity
modes could not be observed in experiments due to the heating
effect, which is not taken into account in the present model. The
temperature dependence of the conductivity $\beta$ is necessary in
order to simulate the overall shape of IV characteristics in the
experiments especially at high currents.

In contrast, the plasma solution without static kinks (crossover to
the McCumber state without radiation) cannot show the sharp peak in
radiation energy when voltage is tuned, and there is no current step
because the plasma do not couple to the cavity modes. On the other
hand, the radiation from the soliton state does not satisfy the ac
Josephson relation, and thus it cannot explain the recent
experiments.

In conclusion, based on computer simulations and theoretical
analysis we reveal the existence of a new dynamic state in
inductively coupled intrinsic Josephson junctions. In this state,
the plasma oscillation is uniform along the $c$ axis with the
fundamental frequency satisfying the ac Josephson relation, while
static phase kinks of $(2m+1)\pi$ are stacked periodically. The
states manifest themselves in the IV characteristics as current
steps at all cavity modes. This state supports quite strong
radiations. The recent experiments on THz radiations from BSCCO
single crystals can be interpreted in terms of this state in a
consistent way .

The authors thank U. Welp, K. Kadowaki, M. Tachiki, L. Bulaevskii,
A. Koshelev and N. Pederson for discussions. Calculations were
performed on SR11000 (HITACHI) in NIMS. This work was supported by
WPI Initiative on Materials Nanoarchitronics, MEXT, Japan,
CREST-JST, Japan and ITSNEM of CAS.


\end{document}